# Post-pandemic mobility patterns in London

Roberto Murcio[1,2] Nilufer Sari Aslam[1], Joana Barros[1]

1 Geography, Birkbeck, London University

2 Centre for Advanced Spatial Analysis, University College London

**Abstract**

Understanding human mobility is crucial for urban and transport studies in cities. People's daily activities provide valuable insight, such as where people live, work, shop, leisure or eat during midday or after-work hours. However, such activities are changed due to travel behaviours after COVID-19 in cities. This study examines the mobility patterns captured from mobile phone apps to explore the behavioural patterns established since the COVID-19 lockdowns triggered a series of changes in urban environments.

## 1   Introduction

The COVID-19 pandemic has had lasting effects on urban mobility behaviour, particularly among office workers, who have since adopted more flexible working patterns. According to ONS [ONS, 2023], 40% of London residents work in hybrid mode, the highest reported level in Great Britain. A recent survey with central London workers found that half of respondents work in the office at least three days a week, with two days a week being the most popular working pattern (Swinney, 2023). Much has been said about the impact of the pandemic on the urban environment, which was felt across the housing market, offices vacancy rates (Local Data Company, 2023), decreased footfall around businesses, and expenditure habits - to name a few. According to Transport for London (TfL, 2022)., as of October 2022, traffic levels in London were nearly back to pre-pandemic levels at 94 per cent and average daily demand for bus and underground services, respectively, 84 and 82 per cent Interestingly, there is also evidence that travel levels are lower in London's central areas (Batty, 2023; TfL, 2023) in comparison to the overall Greater London Authority (GLA). TfL also highlights that, in contrast, active modes such as cycling have shown an increasing trend - a pattern observed in other countries in Europe (Wohner, 2023;).

Much attention has been given to the geographical changes of flexible working since the pandemic, focusing on local areas and a hollowing effect in central areas of larger cities like London (Batty, 2023; Heathcote, 2023). Less debated are the effects of temporal flexibility – as Wohner (2023) pointed out, which means that workers can choose when to work in the office. In London, the effects on temporal change travel behaviour have been noted by TfL (2023), with reported changes to activity levels throughout the week, albeit not across all travel modes. TfL reported no changes in road traffic and bus demands throughout the week compared to pre-pandemic periods. The recovery of rail and underground travel levels was faster on weekends, and there are also noticeable differences across the days of the week. Unlike pre-pandemic patterns, central days - Tuesday to Thursday - are now the busiest days of the week regarding travel levels. TfL's (2023) findings are confirmed by the Centre for Cities' survey, which shows Tuesday to Thursday are the most popular days, while Friday – the busiest Day pre-pandemic according to TfL - is the least popular (Swinney, 2023).

Since the start of the pandemic, there has been a surge of studies on mobility behaviour and much speculation about the future of cities. As we distance ourselves from the pandemic and the threat of new lockdowns, studies on post-pandemic trends emerge. There are questions on whether these new patterns are anything but transitionary, part of a slow recovery period, or this is a new urban normalcy (Batty, 2023). Either way attempts to address the ill effects of the pandemic on city centres and businesses are underway (Local Government, 2022), which, if successful, will affect the current dynamics. Hence, it is fair to assume that further change is coming. Nevertheless, it is paramount that we gain a better understanding of the current state of affairs – be it to document the transition, enable effective planning for current demands, or use the knowledge to attempt to change it.

The present study contributes to understanding post-pandemic mobility behaviour, focusing on London and using the GLA region as a case study. Travel and mobility behaviour studies have traditionally relied on survey data in the form of travel surveys or diaries, albeit rich in details at individual level data, imposed constraints on sample size and study areas. More recently, data collection has been facilitated by popularising smartphones with GPS capability, which enabled researchers to gather detailed individual mobility data. Using tailored-made apps, such studies have taken advantage of gathering unbiased travel trajectories over longer periods and collecting information on individuals' socio-economic characteristics. Some studies also complemented this information with questionnaires and interviews, gaining further insights into travel behaviour motives.

Studies of mobility using alternative sources such as smart-card data and mobile phone records have recently proliferated. Big Data has enabled large-scale case studies and provided unprecedented insights into human mobility at a detailed geographical and temporal level. The main challenges for using such datasets are related to the preservation of the anonymity of individuals, which results in the absence of socio-economic characteristics, and/or the aggregation of data in spatial and temporal units, meaning that individual trajectories cannot be identified from the dataset. Whilst useful for understanding mobility patterns, the absence of individual information means these datasets cannot replace travel surveys and diaries(Wolf et al., 2001; Deng and Ji, 2010; Yazdizadeh et al., 2019)

The present study contributes to methodological efforts for inferring information on individuals from aggregate data. Using a spatial and temporally aggregated dataset from mobile phone apps usage for a month for the GLA area, we create artificial travel diaries that allow us to analyse urban mobility while avoiding the use of sensitive data. These travel diaries capture primary activities (Home-to-Work commute) and secondary activities (school run, groceries, leisure, etc.), as shown in Figure 1. Our initial results showed that Travel to work, shop and amenities trips have changed since the COVID-19 pandemic. The traditional Monday to Friday, 9-5 work pattern has become a work-from-home and fewer commute trips. In line with TfL's (2023) and Swinney (2023), our analysis confirms the empirical observation that Fridays are no longer the most popular Day. Our findings set Thursday as the primary commute day in the new normality.

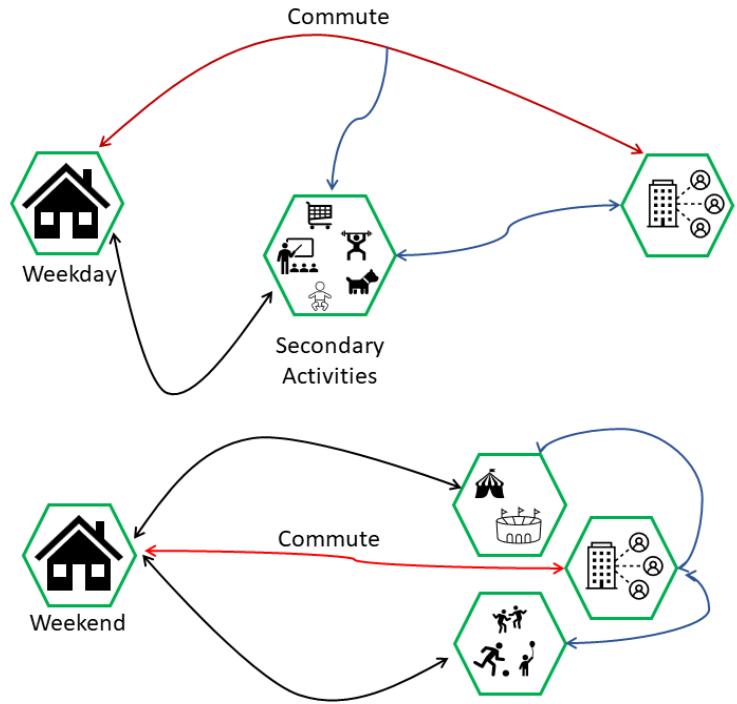

Figure 1. Human mobility in cities. Red lines represent primary activities (i.e. home to work commute); Black lines are secondary activities from home, and blue lines are work/commute travel to secondary activities. Our travel diaries are defined at the Hexagon level. Secondary activities can happen from home or in the middle of the commute. The home-to-work commute (primary activity) is not restricted to weekdays. A large sector of the population works on weekends and similarly does their secondary activities to those on weekdays.

In what follows, we detail the study area and data used for this research under section 2, introduce the proposed methodology for creating artificial travel diaries in section 3, and present our results in section 4. We discuss the overall spatiotemporal patterns of mobility patterns analysed from aggregate data, which are further elaborated based on preliminary results obtained from a sample of individual travel diaries. We conclude the article with a discussion on both the empirical and methodological contributions of the study as well as the next steps of our project.

## 2 Data and study area

### 2.1 Mobile phone data

The mobile data, provided by the company Locomizer (locomizer.com), is sourced from various apps that represent various user types to capture representative and robust coordinate data per unique device that best reflects the population. The data is collected via mobile smartphone applications and opted into by users. If the device has a specific app installed and location data collection is enabled, this data will be accessible. This work's data is from 119,406 Uber hexagons [cite] in the Great London Area (GLA). Each hexagon has an area of $0.015.\,\mathrm{km}^2$ ($0.0007\,\mathrm{km}^2$ squared for the pentagon area).

For this work, Locomizer provided two products built with the above data: 1)Origin-Destination flows and 2)Footfall counts. Both datasets present an aggregated number of users at the hexagon level, with a temporal variation of nine categories:

Table 1. Time intervals

| TIME INTERVAL | HOURS |
|---|---|
| 1 | 04.00-07.59 |
| 2 | 08.00-09.00 |
| 3 | 10.00-11.59 |
| 4 | 12.00-13.59 |
| 5 | 14.00-15.59 |
| 6 | 16.00-17.59 |
| 7 | 18.00-19.59 |
| 8 | 20.00-23.59 |
| 9 | 00.00-23.59 |

## 2.2 Origin-Destination (OD)

This dataset is a typical OD matrix with the registered flows from one hexagon to another at each time interval during June 2022. The datasets have two types of users: workers and All (residents + transient) users. Resident users are obtained from an area where a user has the highest dwell time during the night hours (00.00-06.00 with a heavier weighting for 02.00-04.00), called the common evening locations as home locations/residential areas. Note that international visitors are excluded from this classification. Workers are obtained from an area where a user has the highest dwell time during working hours. Such common daytime locations are called work. Transient users are obtained from an area that does not live or work, consisting of local and international visitors. Note that international users can be in this group even though they do not live in a residential area. Table 2 shows the basic statistics for the workers and All data.

Table 2: OD – Descriptive statistics from OD from workers and all datasets. Quantities exclude zeros, i.e., OD pairs with zero trips.

| All flows | | | | |
|---|---|---|---|---|
| Count | Mean | Std | Min | Max |
| 131,718,666 | 32.18 | 22.65 | 22 | 1750 |
| Worker flows | | | | |
| Count | Mean | Std | Min | Max |
| 14,208,621 | 51.26 | 41.51 | 22 | 875 |

To understand the representation of this data, we aggregated the whole month of flows by OD pair. The mean of the aggregated flows is 133, and as expected at this scale [cite], the vast majority (75%) of the OD pairs have fewer trips than this mean (All and work).

In terms of the spatial distribution, looking at the aggregated Origins/Destinations for both work and All datasets, we found expected mobility patterns, as we can see in Figure 1. For the Origin (Fig 1. A), All data, all the major green areas in London have practically zero counts, i.e., no trips start in a park. Also, we have low counts in The City and Canary Warf areas, which are the financial hubs with a small number of residents, so trips in from these areas start to appear later in the Day, affecting the aggregated number of flows in a month. The All Destination spatial distribution (Fig 1. B) presents a different picture. People seem to follow the transport and road networks with striking accuracy. Regardless of the Origin, people use the provided infrastructure to move around.

To some extent, this map mirrors Fig 1. A, having high Origin counts vs low Destinations flows, and vice versa. The prime examples of this dichotomy are The City/Canary Wharf areas.

Regarding the work dataset, the Origin distribution is similar to the All, only changing in magnitude.

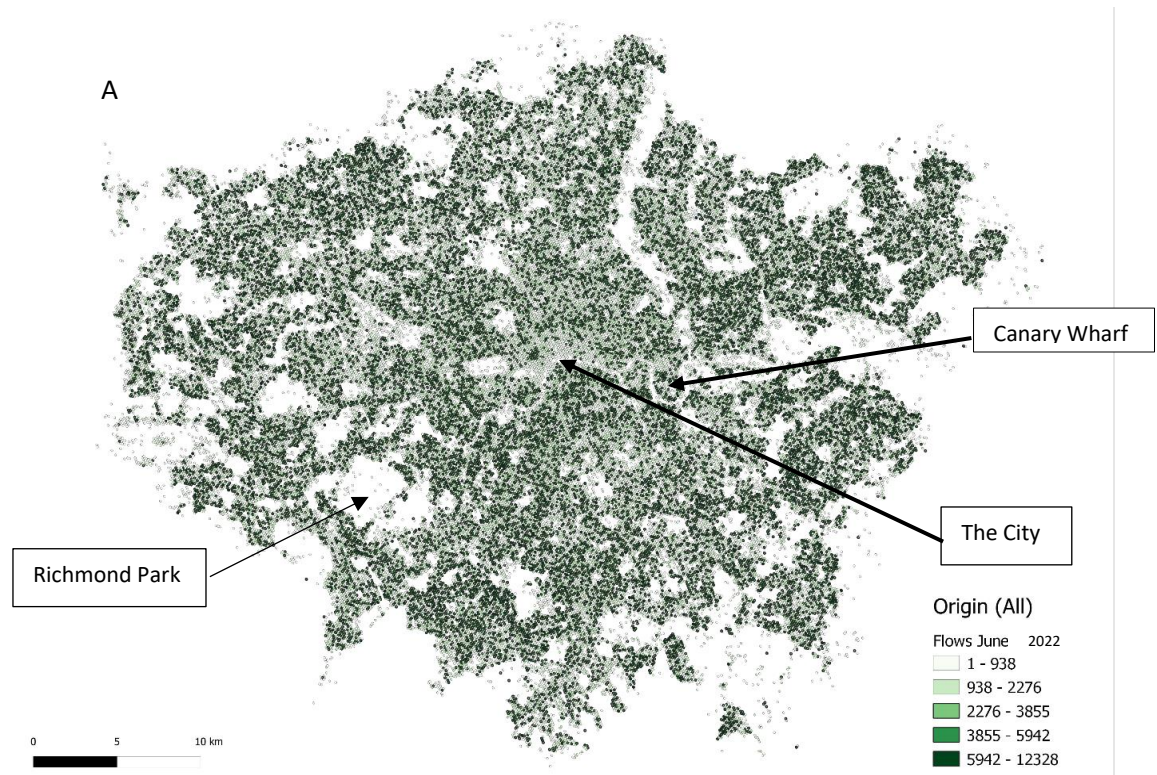

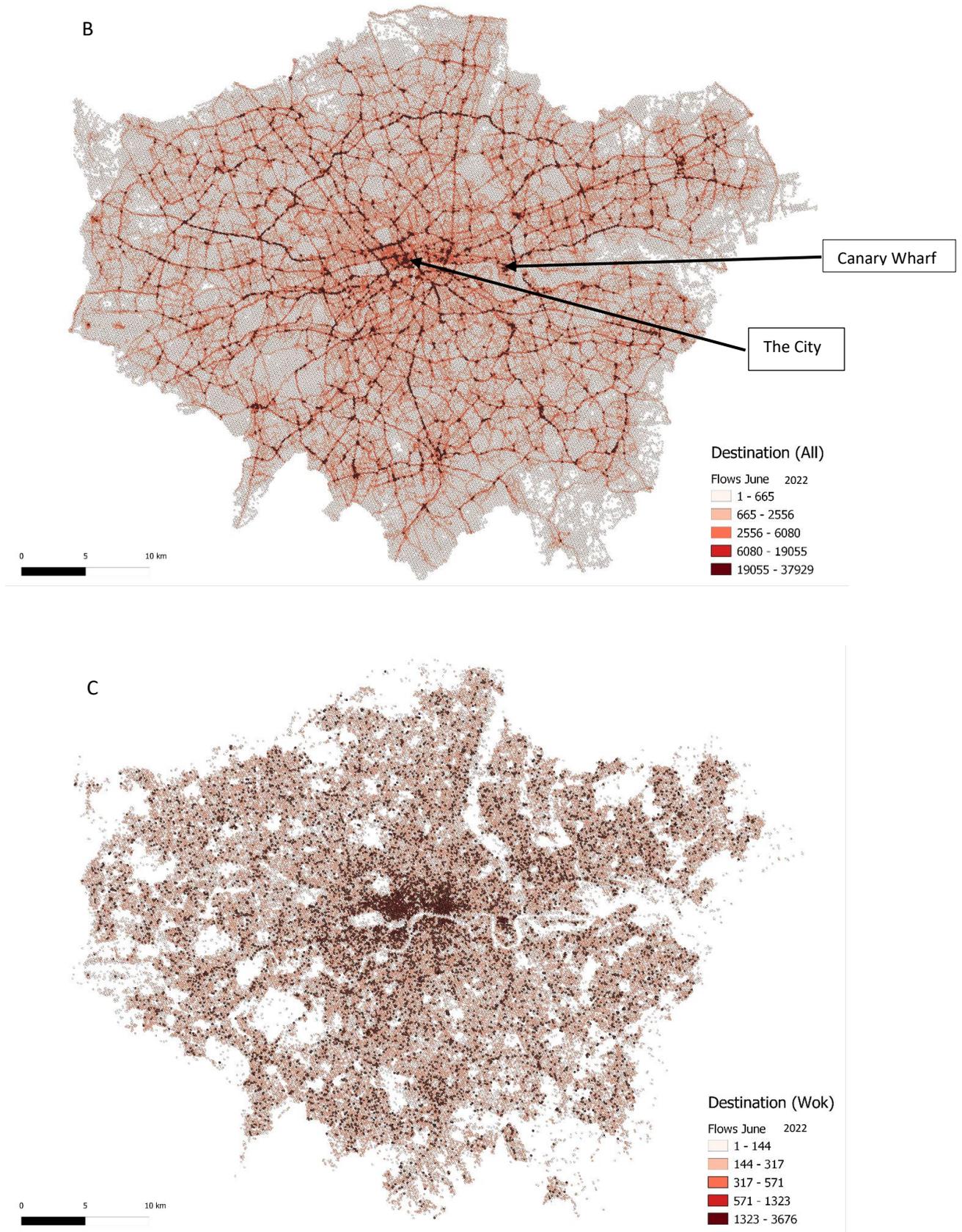

Figure 1. Aggregated flows for June 2023. A. Origin All. We can observe the large empty areas with no origins for the micro-flows, which correspond to green areas and parks. We can also see that the areas around the City and Heathrow have low counts, which makes sense as those are fundamentally destination zones. B. Destination All. The highest flows appear to follow the primary and secondary streets in London. Surprisingly, Heathrow was not a major destination in the aggregated counts. C. Destination Workers. The difference with the Origin All is clear: green areas are empty, and The City is where hexagons have the largest share of destination flows. Spatial distribution of the Origin flows in the work dataset. As expected, The City, Canary Wharf and the areas around central London hold the bulk of the work destinations

Finally, we explored the temporal representation of the OD data. We found that there are no missing day/time entries for any of the OD pairs in both All and work datasets. With this high temporal granularity, we can explore the changes in the type of activities during a single day and start detecting patterns between similar socio-economic areas. As an example (Figure 2), we follow the change in user counts around two typical working locations, Waterloo Station and The City of London square mile

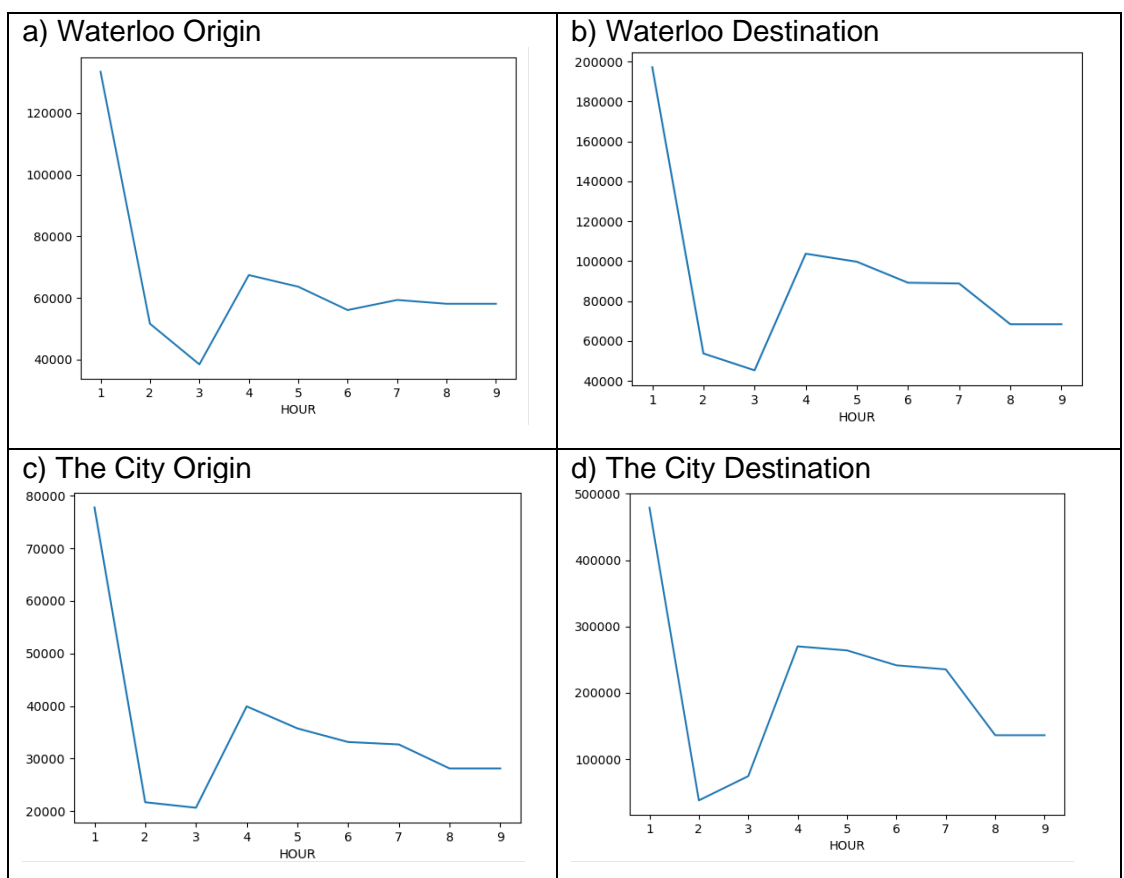

Figure 2. Aggregated temporal variation. a) Waterloo station area. Origin and Destination counts by hour show the same trend, a high peak at time 1 (04.00-07.59) followed by fewer users throughout the Day. Besides the evident difference in magnitude (200,00 users as a destination while only 14,000 as Origin for time 1), we observed that Waterloo, as an Origin, keeps a steady trend around 6,000 from hour 6. In contrast, as a destination, the counts decrease from hour 7, showing fewer people arrive at this station by the end of the Day. b) The city Area is similar in shape to Waterloo (peak in the morning and then a significant decrease) and a peak at time 4 (12.00-13.59). Moreover, as Waterloo, the difference in magnitude is substantial, showing that these two locations are primarily Destinations.

## 2.3 Footfall Data (FF)

This dataset provides the key footfall metrics where people spend time at the hexagon level. The metrics are built using the mobile trace data generated by a representative panel of anonymous users. This includes the observed users who spent time at the point during a specific period. For the FF data, the type of users is as in the OD data, but with the resident and transient users disaggregated.

# 3 Methodology

## 3.1 Typical Home-Work detection

We follow a simple approach to detect Home-Work locations in the OD work data. Every Day, we selected all records at times T1 and T2, which destinations are their Origins at T6, T7. To avoid capturing casual connections, we look if the above procedure repeats at least ten days during June to label an OD pair as Home-Work.

We are looking for those hexagons that create a typical "9-to-5" work travel diary without any other secondary activity in the middle. For example, Figure 3 shows a mock-up OD configuration between Hexagons. H1 is the Origin for all Hexagons, and it is the Home location for hexagons H2, H4 and H5, as it fulfils the pre-established conditions. H1 is not home location for H3, as although they are connected at T1, H3 is the Origin of H1 at T3

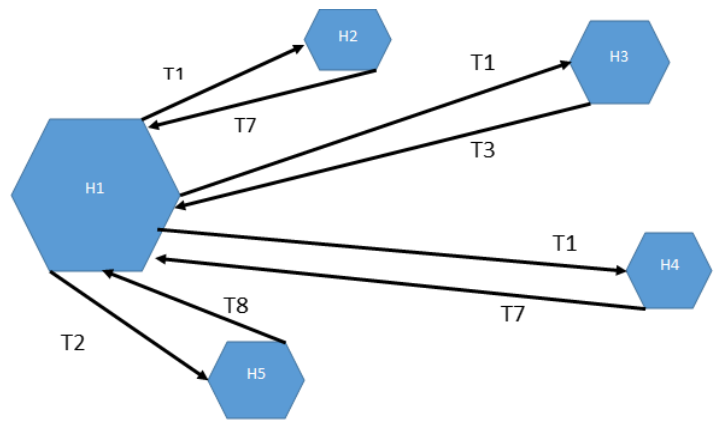

Figure 3. Mock-up configuration for an OD pattern among five hexagons.

After building our work-home OD matrix *M*, we create travel work diaries following the algorithm explained in section 3.2.

As an example of the OD data, Figure 4 shows the flows from the top ten origin and destination hexagons regarding the number of users.

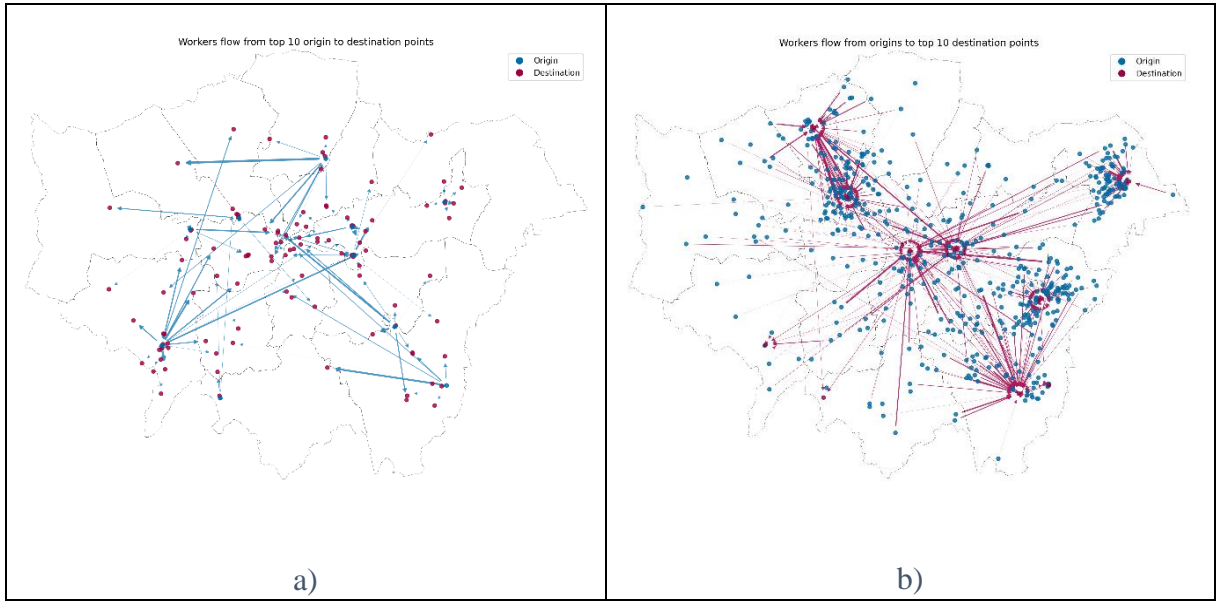

Figure 4: a) Workers flow from the top 10 Origins. Large origins are biased towards certain demographics (Kingston Upon Thames, Bromley and Walthamstow) b) Workers flow to the top 10 destinations. As expected, Westminster, The City and Canary Wharf emerge in this pattern, but also we can observe some interesting micro hubs around Ilford and not-so-obvious locations.

## 3.2 Travel work diaries algorithm

Our algorithm tracks all the work morning destinations and finds where users in these hexagons move from throughout the Day. We defined four temporal regimes in our weekday data (Table 3) to reflect the activities that may occur during a regular weekday.

Algorithm A1
1. From M, select all flows at T1/T2. This set is M1
2. Select all Ti flows with Origin all Destinations in Mi-1 from M. This is Mi, with i=2..9
3. Using a classic Eclat mining algorithm, obtain the flow patterns in Mj, j=1..9.
4. Once the patterns are identified, we enrich each zone with the POI and FF data.

*Table 2 Temporal regimes*

| Morning peak | | Lunch Break / School run / Shopping | | | Evening peak / Leisure / Shopping | | Night out/ Night workers | |
|---|---|---|---|---|---|---|---|---|
| T1 | T2 | T3 | T4 | T5 | T6 | T7 | T8 | T9 |

## 4 Results

### 4.1 Aggregated figures

Looking at the daily number of users at Origin and Destination locations, we discovered a clear shift in behaviour post-Pandemic. Figure 5 shows how Thursdays (day 4) is the busiest Day in the week at Origins and Destinations. At both locations, we have identical behaviour. Mondays are the quietest Days of the working week, and the number of users slowly grows during Tuesdays and Wednesdays to explode on Thursdays, returning to Monday levels for the whole weekend.

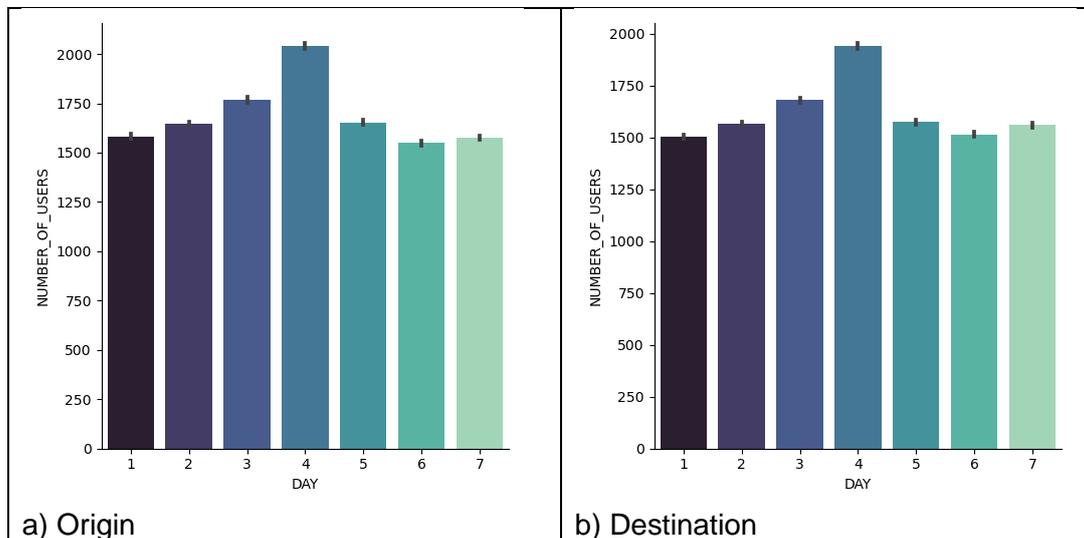

Figure 5. Box plots by Day. Days are enumerated from 1-Monday, 2-Tuesday until 7-Sunday. In a) and b), day 4-Thursdays is the busiest Day in the week, while Fridays are now at the same level as Mondays. Both worker locations behave similarly, with a) having similar weekends and b) with Saturdays with a smaller number of flows.

Furthermore, we can observe that the difference Thursday-Weekend is not only in magnitude but is focalised, in the case of the work dataset, to traditional work locations in London.

Getting the difference between the number of flows between Thursday and Sunday (Figure 6) provides further evidence about the importance of Thursday in the new normality. We can observe how The City and Canary Wharf greatly differ between Thursday and Sunday counts.

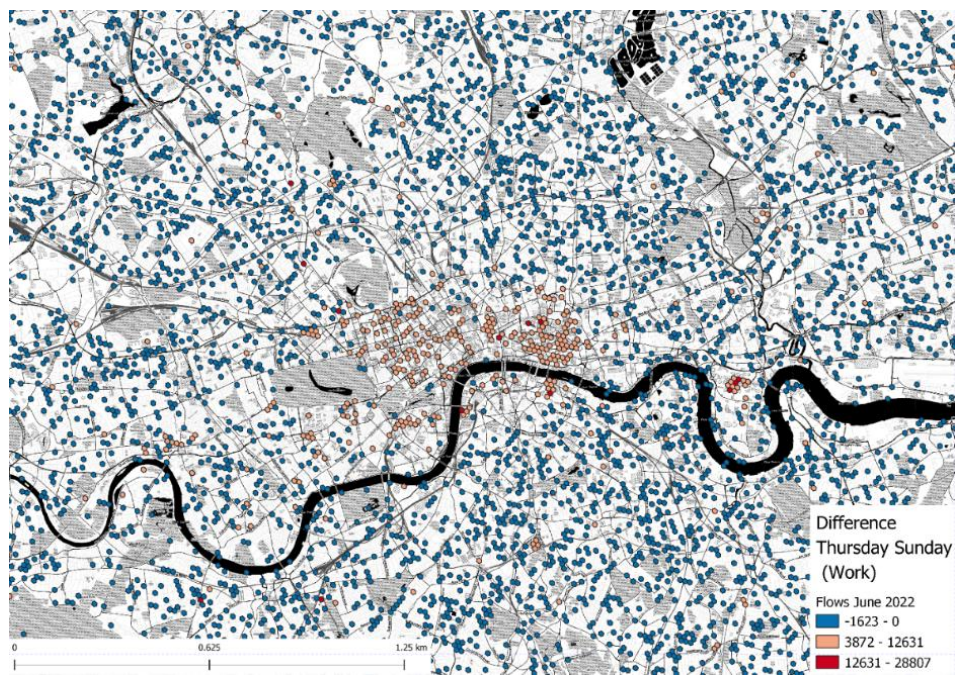

Figure 6. Difference between Thursday and Sunday. The City and Canary Wharf lost all the flows on Sunday as expected, being typical working locations, while the blue areas gained

flows on Sunday. Interestingly, most hexagons gain flows on Sunday as more people travel for leisure to different areas.

## 4.2 Travel diaries workers

Following algorithm A1, we obtain the following example for the Cricklewood area in Northwest London. Cricklewood is a typical working-class location, with some areas with high indices of Multiple Deprivation. Figure 9 shows the data obtained for Hexagon 8a195da43687fff (51.5565,-0.2173) for all Thursdays in June 2023.

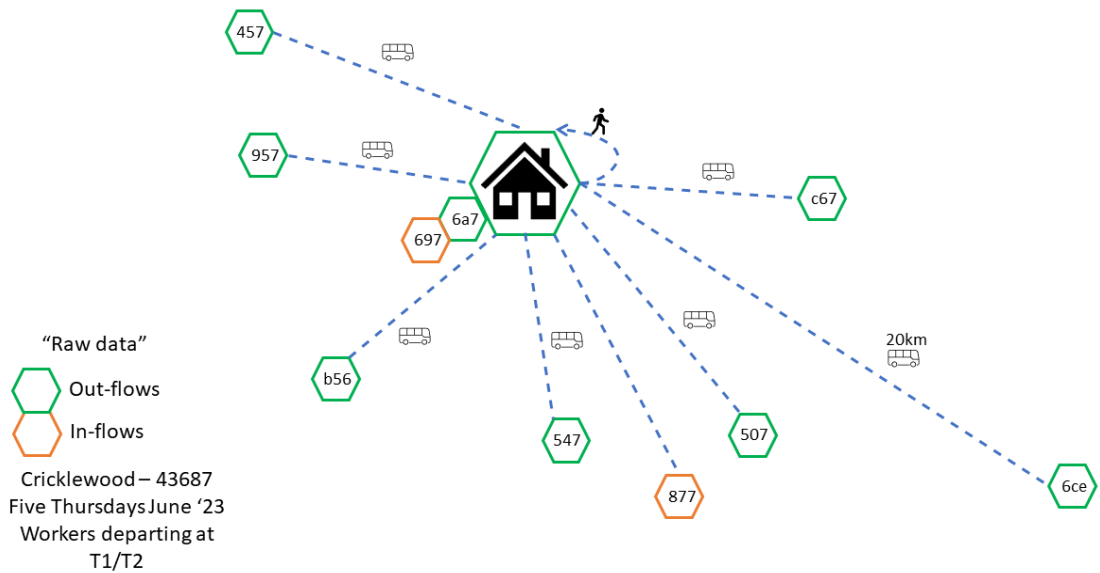

Figure 7. This schematic shows a hexagon (8a195da43687fff) at the centre as the origin point at times T1/T2. It provides flows to ten hexagons at morning peak, some as far as 6ce, 20km apart. At the centre, the self-loop represents the intraflows. Areas 6a7 and 697 are the nearest neighbourhood hexagons that are at less than 100 m distance. The bus icon represents that those hexagons are not easily reachable by foot. Two hexagons, 697 and 877, provide inflows to the 8a195da43687fff.

After applying our Eclat algorithm and enriching our results with POIs, we obtain the following pattern in Figure 8:

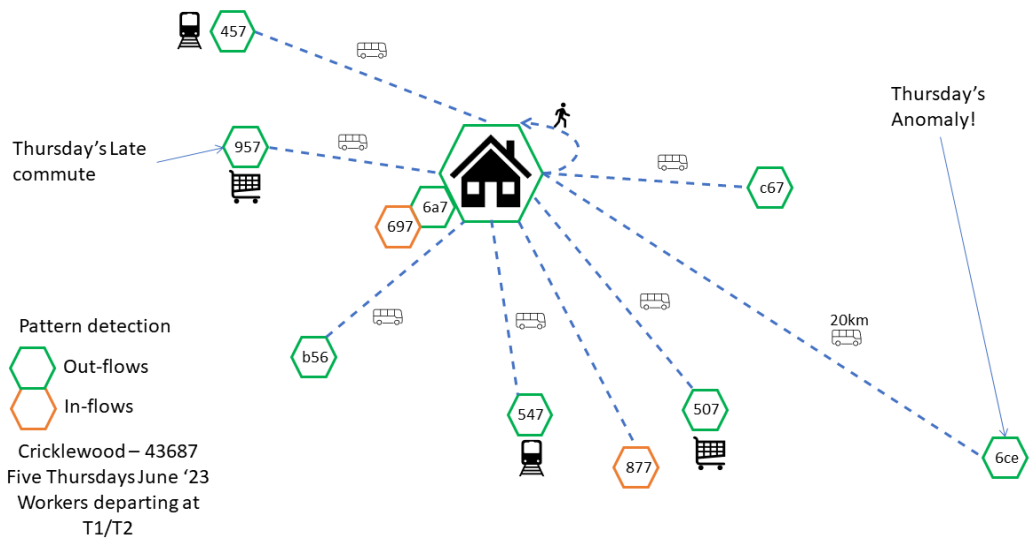

Figure 8. Patterns obtained from the Eclat algorithm. All the OD flows from the raw data prevailed. However, after comparing it with Mondays, June 2023, we discovered that flows to 6ce and 957 (with a large depot store) are particular for Thursdays. Hexagons 457 and 547 contain important railway stations (Hendon and West Hampstead). Moreover, hexagon 507 corresponds to one of the largest departmental stores in Oxford Street, London.

The travel diaries built for hexagon 8a195da43687fff show the evolution through the temporal regimes defined in Table 2. The bulk of the flows corresponds to the intraflows, i.e., flows starting and ending at the same hexagon. Some locations disappear as the day advances, but no new hexagons appear, indicating the stability of this particular location. figure

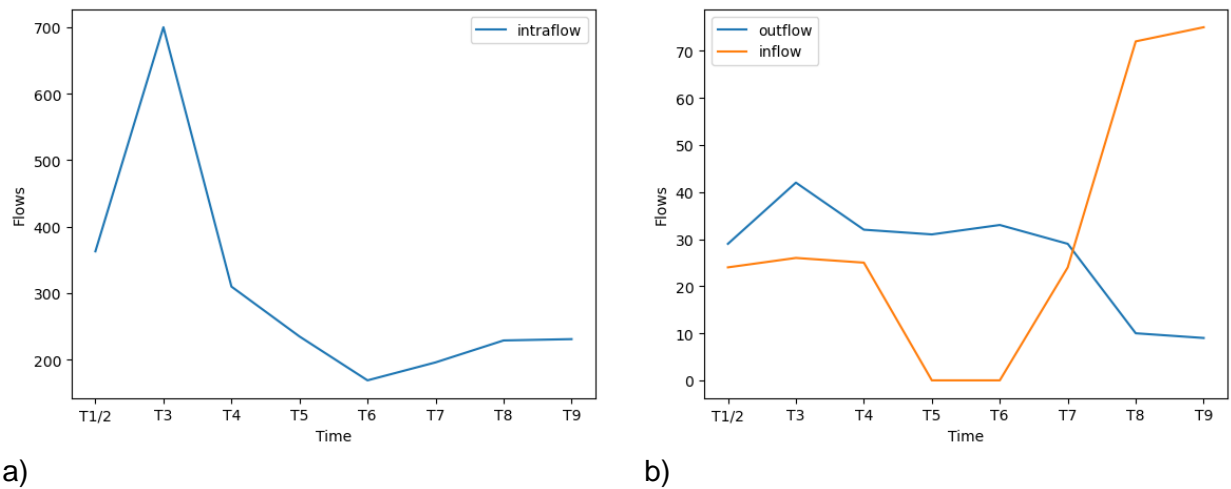

a) b)

Figure 9. Flow counts for Cricklewood Hexagon. The intraflows a) dominated the counts, with massive movements before lunchtime. This is characteristic of commercial areas like this one. We can observe at b) that the inflow has an important increase from T7 when people return home. These counts at the evening peak (T6) are smaller than for Mondays.

## 5  Conclusion